\documentstyle[epsfig]{mn}
\newif\ifAMStwofonts

\ifoldfss
  \ifCUPmtlplainloaded \else
    \NewTextAlphabet{textbfit} {cmbxti10} {}
    \NewTextAlphabet{textbfss} {cmssbx10} {}
    \NewMathAlphabet{mathbfit} {cmbxti10} {} 
    \NewMathAlphabet{mathbfss} {cmssbx10} {} 
  \fi
  \ifAMStwofonts
    \ifCUPmtlplainloaded \else
      \NewSymbolFont{upmath} {eurm10}
      \NewSymbolFont{AMSa} {msam10}
      \NewMathSymbol{\upi}     {0}{upmath}{19}
      \NewMathSymbol{\umu}     {0}{upmath}{16}
      \NewMathSymbol{\upartial}{0}{upmath}{40}
      \NewMathSymbol{\leqslant}{3}{AMSa}{36}
      \NewMathSymbol{\geqslant}{3}{AMSa}{3E}

       \let\ge=\geqslant
    \fi
  \fi
\fi 

\ifnfssone
  \newmathalphabet{\mathit}
  \addtoversion{normal}{\mathit}{cmr}{m}{it}
  \addtoversion{bold}{\mathit}{cmr}{bx}{it}
  \newmathalphabet{\mathbfit} 
  \addtoversion{normal}{\mathbfit}{cmr}{bx}{it}
  \addtoversion{bold}{\mathbfit}{cmr}{bx}{it}
  \newmathalphabet{\mathbfss} 
  \addtoversion{normal}{\mathbfss}{cmss}{bx}{n}
  \addtoversion{bold}{\mathbfss}{cmss}{bx}{n}
  \ifAMStwofonts
    \ifCUPmtlplainloaded \else
      \UseAMStwoboldmath
      \new@mathgroup\upmath@group
      \define@mathgroup\mv@normal\upmath@group{eur}{m}{n}
      \define@mathgroup\mv@bold\upmath@group{eur}{b}{n}
      \edef\UPM{\hexnumber\upmath@group}
      \new@mathgroup\amsa@group
      \define@mathgroup\mv@normal\amsa@group{msa}{m}{n}
      \define@mathgroup\mv@bold\amsa@group{msa}{m}{n}
      \edef\AMSa{\hexnumber\amsa@group}
      \makeatother
      \mathchardef\upi="0\UPM19
      \mathchardef\umu="0\UPM16
      \mathchardef\upartial="0\UPM40
      \mathchardef\leqslant="3\AMSa36
      \mathchardef\geqslant="3\AMSa3E

       \let\ge=\geqslant
    \fi
  \fi
\fi 

\ifnfsstwo
  \DeclareMathAlphabet{\mathbfit}{OT1}{cmr}{bx}{it}
  \SetMathAlphabet\mathbfit{bold}{OT1}{cmr}{bx}{it}
  \DeclareMathAlphabet{\mathbfss}{OT1}{cmss}{bx}{n}
  \SetMathAlphabet\mathbfss{bold}{OT1}{cmss}{bx}{n}
  \ifAMStwofonts
    \ifCUPmtlplainloaded \else
      \DeclareSymbolFont{UPM}{U}{eur}{m}{n}
      \SetSymbolFont{UPM}{bold}{U}{eur}{b}{n}
      \DeclareSymbolFont{AMSa}{U}{msa}{m}{n}
      \DeclareMathSymbol{\upi}{0}{UPM}{"19}
      \DeclareMathSymbol{\umu}{0}{UPM}{"16}
      \DeclareMathSymbol{\upartial}{0}{UPM}{"40}
      \DeclareMathSymbol{\leqslant}{3}{AMSa}{"36}
      \DeclareMathSymbol{\geqslant}{3}{AMSa}{"3E}

       \let\ge=\geqslant
    \fi
  \fi
\fi 

\ifCUPmtlplainloaded \else
  \ifAMStwofonts \else 
    \def\upi{\pi}
    \def\umu{\mu}
    \def\upartial{\partial}
  \fi
\fi

\title[Expectations for a Galaxy Cluster Sky Survey with 
AMI]{Surveying the Sky with the Arcminute MicroKelvin Imager: 
Expected Constraints on Galaxy Cluster Evolution and Cosmology}

\author[Kneissl et al.]
  {R\"udiger~Kneissl$^1$, Michael E. Jones$^1$, Richard Saunders$^1$, 
Vincent R. Eke$^{2,3}$, \newauthor 
Anthony N. Lasenby$^1$, Keith Grainge$^1$ and Garret Cotter$^1$ \\ 
$^1$Astrophysics, Cavendish Laboratory, Cambridge University, 
Madingley Road, Cambridge CB3 0HE, UK \\ 
$^2$Institute of Astronomy, Cambridge University, 
Madingley Road, Cambridge CB3 0HA, UK\\ 
$^3$Steward Observatory, 933 N Cherry Ave, Tucson AZ 85721, USA}
\date{Accepted Received In original form}
\pagerange{\pageref{firstpage}--\pageref{lastpage}}
\pubyear{2001}

\def\LaTeX{L\kern-.36em\raise.3ex\hbox{a}\kern-.15em
    T\kern-.1667em\lower.7ex\hbox{E}\kern-.125emX}

\def\deg{^\circ}

\begin{document}

\label{firstpage}

\maketitle

\begin{abstract}

We discuss prospects for cluster detection via the Sunyaev-Zel'dovich (SZ)
effect in a blank field survey with the proposed interferometer array, 
the Arcminute MicroKelvin Imager (AMI). 
Clusters of galaxies selected in the SZ effect probe cosmology and 
structure formation with little observational bias, because the effect 
directly measures integrated gas pressure, and does so independently of 
cluster redshift. 

We use hydrodynamical simulations in combination with the Press-Schechter 
expression to simulate SZ cluster sky maps. These are used with simulations 
of the observation process to gauge the expected SZ cluster counts. 
Even with a very conservative choice of parameters 
we find that AMI will discover at least several tens of clusters every year 
with $M_{\rm tot} \ge 10^{14} M_\odot$; 
the numbers depend on factors such as the mean matter density, the 
density fluctuation power spectrum and cluster gas evolution. The AMI survey 
itself can distinguish between these to some degree, and parameter 
degeneracies are largely eliminated given optical and X-ray 
follow-up of these clusters; this will also permit direct investigation 
of cluster physics and what drives the evolution. 

\end{abstract}


\section{Introduction}

Clusters of galaxies are the most massive collapsed objects in the
universe. They are ideal probes of structure formation, both in the linear
clustering regime, which is controlled by factors such as the cosmological 
density parameter $\Omega_{0}$ and the 
density fluctuation power spectrum, and later when the individual clusters
grow non-linearly by merging, shocks, and gradual virialisation.  It is 
becoming
clear that there is a population of clusters $z\ge1$ (see e.g.  Fabian et
al. 2001, Cotter et al., Joy et al.), but the numbers of these remain
unclear. It is essential to search systematically for these clusters in order
to understand the evolution of structure in the universe.

To make full use of these probes of cosmology, an unbiased cluster survey is
needed. Optical and X-ray surveys suffer various biases with respect to
cluster properties and redshift.  Projection effects, confusion with
background objects, surface brightness dimming with redshift, and a bias to
mass concentration, all hamper both optical and X-ray surveys, although X-ray
luminosity turns out to be a reasonably good indicator of the total mass in
many clusters.  However, detecting clusters via the scattering of CMB photons
by cluster gas (the Sunyaev--Zel'dovich, SZ, effect, Sunyaev \& Zel'dovich
1972) minimises such biases.

First, the measured quantity, integrated gas pressure, gas mass times gas 
temperature, or total thermal energy is closely related to the cluster mass 
(e.g. Bartlett and Silk 1994; Barbosa et al. 1996; Eke et al. 1996).  
Second, critically,
because the SZ effect is a scattering process, the effect is independent of
cluster redshift. Surveys using the SZ effect therefore offer the means to
find clusters by mass, and at all redshifts.

Detection of the SZ effect in rich X-ray-selected clusters has become routine
(see Birkinshaw 1999 for a review) Although the use of microwave cluster
searches for cosmological studies has be advocated for a long time (Korolev,
Sunyaev \& Yakubtsev 1986; Bond \& Myers 1991; Bartlett \& Silk 1994;
Markevitch et al. 1994; Barbosa et al. 1996; Eke et al. 1996; Colafrancesco et
al. 1997; da Silva et al. 2000) only very recently has the technology and
expertise become available to build sufficiently sensitive instruments that
can survey the sky and are expected to find a substantial number of clusters
in a reasonable surveying time.

In this paper we will explore the expected results from an SZ survey using the
characteristics of the proposed Arcminute MicroKelvin Imager (AMI). Other
related experiments are proposed or planned (SZA: Holder et al. 2000, AMiBA:
Lo et al.).  Predictions for the results of this kind of survey exist now also
in the literature (e.g. Holder et al. 2000, Bartlett). The work presented here
(cf. Kneissl 2000 for an earlier presentation) is different in two main ways:

\begin{itemize} 

\item We have tried to be as conservative as possible about what an AMI survey
will find, given what is already known about structure formation.  For
example, Holder et al. (entirely reasonably) expect to find larger numbers of
clusters than we report here; however, this difference is only due to their
different choice of model parameters: f$_g$ (gas fraction) higher by 50\% and
$\sigma_8$ (rms density fluctuations on 8 Mpc scales) higher by 20\%. This
change in the parameters leads to a number of 
clusters that is higher by a factor of ten. 
This large difference in expected source counts for a plausible
variation in input parameters clearly demonstrates the importance of
observations which will have an error of the order of only about 10\%.  Since
the uncertainties in the model parameters entering the SZ cluster simulation
are large, we take a cautious approach and demonstrate that $\Omega_{0}$ can
be sensibly constrained even with a pessimistic parameter choice.  With more
optimistic assumptions for the model parameters, in particular the gas
fraction and the local cluster abundance, more clusters are expected, and
higher precision in constraining parameters can be achieved.  Constraints on a
cosmological constant or even quintessence (Wang and Steinhardt 1998, Haiman
et al.) are possible in principle and are certainly a fascinating
possibility. Also deviations from the expected gravitational evolution allows
us to identify non-Gaussian initial conditions (e.g. Matarrese et al. 2000).

\item We model what {\it interferometers} will detect, given their uv
coverage, their errors, and given the ubiquitous existence of contaminating
radio sources.

\end{itemize} 

AMI is described and a brief overview over its other science goals is given in
Section \ref{sec:ami}. To gauge the ability of AMI to detect clusters, we 
simulate the SZ cluster sky using the Press-Schechter expression and
individual cluster templates from hydrodynamical simulations, described in
Section \ref{sec:model}. The expected cluster counts and instrument
sensitivities are considered in Section \ref{sec:clusters}, and in Section
\ref{sec:parameters} we discuss the exciting prospects of using AMI results to
determine the model parameters, and follow-up observations, both pointed SZ
observations and in other wavebands. We argue that detection in the SZ effect
is vital for selection, and follow-up in the optical and X-ray bands is
necessary to break degeneracies between the model parameters.

\section{The Arcminute MicroKelvin Imager}
\label{sec:ami}

It has not yet proved possible to conduct an effective blind SZ survey
because of the limited sensitivity and field of view of current
telescopes. A very rich cluster produces a perturbation of less than
1~mK on the CMB, over an angular size (for a moderate-to-high redshift
cluster) of a few arcminutes. A small number of groups have been able
to detect successfully the SZ effect in X-ray and optically selected
clusters (e.g. Birkinshaw, Gull \& Hardebeck 1984; Jones et al. 1993; 
Carlstrom, Joy \& Grego 1996; Holzapfel et al. 1997; Myers et al. 1997). 
However, existing telescopes do not have sufficient sensitivity over a large
enough field of view to carry out a survey that would usefully
constrain the population of clusters at high redshifts where X-ray and
optical surveys are incomplete.

As an example, our present programme uses the Ryle Telescope (RT) at
15~GHz to make images of the SZ effect in X-ray selected clusters. The
RT, with its compact array of five 13~m-diameter antennas, was the
first instrument in practice capable of imaging the SZ effect, and
remains one of only a handful of such worldwide. We have obtained
$\sim$10-$\sigma$ detections towards a dozen clusters (see
e.g. \cite{1993MNRAS.265L..57G,1996MNRAS.278L..17G}). However, it
takes about twenty 12-hour observations with the RT to achieve a
5-$\sigma$ detection on a rich (total mass~$\approx 10^{15}{\rm
M_{\odot}}, kT \approx 8$keV) cluster, with a field of view of 0.01
square degrees. Models of structure formation normalised to the local
space density of clusters e.g. \cite{1996MNRAS.282..263E} suggest that
the surface density of such clusters is at most of the order 0.1 per
square degree. Thus to detect a few such clusters with the RT in a
blind survey would take over fifty years.

Given the experience of existing SZ and CMB telescopes, it is now
possible to build an SZ survey telescope with the required
performance. Interferometers have advantages over single antennas for
the kind of measurements we need to make. They are less susceptible to
spillover signals in the sidelobes, as signals originating far from
the main beam are attenuated both by delay error and by having the
wrong fringe rate. Receiver stability is not critical as gain
fluctuations are not correlated between different antennas and are
therefore not coherently detected by the correlator. They can be very
insensitive to most atmospheric emission, both by resolving it out
spatially, and by virtue of the fact that it has a very different
fringe rate to the astronomical signal and can be filtered out
temporally \cite{church95,lay2000}. With suitable design,
interferometers can also overcome the ever-present problem of radio
sources contaminating the CMB signal by simultaneously identifying the
sources with high angular resolution and subtracting them from the
short-baseline data.

The design of an SZ survey interferometer has to address two basic
issues:
\begin{itemize}
\item{Good temperature sensitivity is required: we must maintain a
high filled fraction of the synthesised aperture yet have a sufficient
range of baselines to make a good image. Also we must not resolve out
the extended structure of most SZ sources -- their typical angular size
of a few arcminutes requires baselines of a few hundred wavelengths or
less in order to observe most of the flux. We therefore need a 
closely-packed array of small antennas.}
\item{ It is also vital to find and remove the effects of
contaminating foreground emission. In practice on arcminute scales,
the contamination comes from discrete point sources---radio galaxies
and quasars. Their effects can only be removed by mapping the AMI
field of view with higher angular resolution and flux
sensitivity. This requires an array of large antennas with relatively
long baselines.}
\end{itemize}

\begin{figure}
\centerline{\epsfig{file=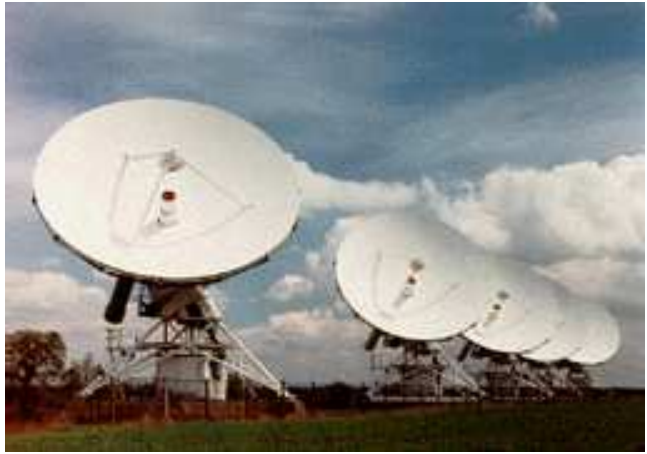,width=8.5cm,angle=0}}
\vspace{0.3cm}
	\centerline{\epsfig{file=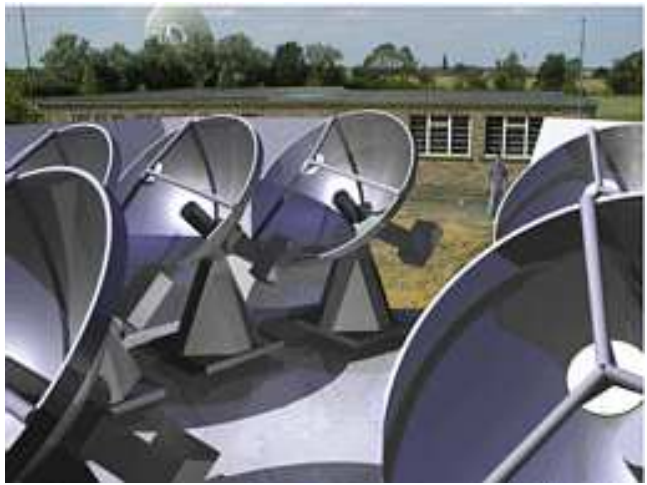,width=8.5cm,angle=0}}
\caption{\small The Arcminute MicroKelvin Imager is planned to 
consist of eight 13-m 
dishes of the Ryle telescope, five of which are visible in the top image, and
an array of 3.7-m antennas, which can be seen in the artist's impression
below.}
\label{fig:amipic}
\end{figure}

The solution to these problems is to use two arrays simultaneously;
larger antennas on longer baselines, providing good flux sensitivity
to point sources, and a compact array of small antennas to provide
sensitivity to the SZ clusters. For AMI, we propose to use all eight
13-m antennas of the RT, with baselines from 18 to 108 m (900 to 5400
$\lambda$ at 15 GHz), plus a new array of ten 3.7-m antennas with
baselines from 4 to 18 m (200 to 900 $\lambda$). The 3.7-m antennas
will be sited inside an earth bank lined with aluminium sheeting to
ensure that sidelobes from the antennas do not terminate on warm
emitting material (see figure~\ref{fig:amipic}).
The filling factor is around 40\% for baselines up to $\sim 2000
\lambda$, giving good temperature sensitivity with a resolution of 1.5
arcmin. The shorter baselines of the RT array provide some sensitivity
to SZ clusters; the remaining longer baselines provide sufficient flux
sensitivity at higher resolution to subtract the radio sources that
will also be present in any field. We will correlate the whole
12--18~GHz band provided by the front-end amplifiers, divided into 8
channels (or 16 in the case of the longer RT baselines) to avoid
losses due to chromatic aberration.

We choose the 12--18~GHz band for several reasons. At a given
resolution and system temperature, any observing frequency within the
Rayleigh-Jeans region of the CMB spectrum ($\nu \ll 217$ GHz) will
give the same sensitivity to temperature structure in the sky. The
relevant questions are then of achievable system temperature and
foreground contamination. Receiver noises are invariably lower at
lower frequency; modern cryogenic amplifiers using pseudomorphic HEMTs
can achieve noise temperatures of 0.5--1 K per GHz of observing
frequency at $\nu < 150$ GHz.  Also, the atmospheric noise
contribution rises with frequency and is more sensitive to the water
vapour content at higher frequencies. On the other hand, the major
foreground contaminant at arcminute resolution is extragalactic radio
sources, whose brightness temperature typically falls as $\nu^{-2}$ to
$\nu^{-3}$. (Note however that Taylor et al., using the RT
at 15 GHz in survey mode, find a significant population of sources
whose flux densities {\em rise} with frequency, i.e. whose brightness
temperatures fall more slowly than $\nu^{-2}$.) Optimum observing
frequency is thus a compromise between system noise and source
contamination. With the RT, a large collecting area is available for
source identification on a site where the atmosphere is very much more
transparent at 12--18 GHz than at the next atmospheric window of
26--36 GHz. System temperatures around 25 K are possible using the
latest amplifiers, and the source confusion problem is quite tractable
using the majority of the RT baselines. Observing solely in the
Rayleigh-Jeans region means of course we are unable to use the
frequency dependence of the SZ effect to separate the thermal and
kinetic SZ effects; however, for the main purpose of making an SZ
survey, sensitivity is the over-riding consideration, and this is best
achieved at lower frequency.

\subsection{Predicted Performance}

The instantaneous field of view, or primary beam, of an interferometer
is given by the Fourier transform of the antenna illumination
function; for typical illumination patterns, the field is well
approximated by a Gaussian with {\sc fwhm} = 1.1$\lambda/D$ where $D$
is the antenna diameter. The flux sensitivity of an array consisting
of $n$~antennas each with effective area~$A$, integrating for a
time~$\tau$, measuring a single polarisation with bandwidth~$\Delta
\nu$ is given by (see e.g. Thomson, Moran \& Swenson 1986)
$$
\Delta S_{\rm rms} = {{2kT_{sys}}\over{\eta A \left( n \left(n-1\right) \Delta
\nu~ \tau \right)^{1/2}}},
$$
where $\eta$ is the system efficiency, $T_{\rm sys}$ is the system
temperature (assumed the same for each antenna), and $k$ is
Boltzmann's constant.  For the two arrays of AMI, this results in flux
sensitivities of $\rm 2.0mJy\, s^{-1/2}$ and $\rm 20mJy \,s^{-1/2}$
over fields of view of 6 arcmin and 21 arcmin respectively. These
figures have been used to generate the sensitivity plots and simulated
observations in the following sections. To give an indication of the equivalent
temperature sensitivity, we assume that the aperture is reasonably
well-filled and use the Rayleigh--Jeans formula. This gives
$$
\Delta T_{\rm rms} = \frac{\lambda^2 \Delta S_{\rm rms}}{2k \Omega},
$$
where $\Omega$ is the synthesised beam area. This corresponds to a
temperature sensitivity of around 8 $\mu$K in a 1.5 arcminute beam in
one month of observation on a single (0.1 square degree) field, though
sensitivity can be concentrated on particular angular scales by choice
of array configuration.

\subsection{Source subtraction}

We now show that AMI can successfully subtract the radio sources that
are the main contaminating signal. The amount of collecting area that
has to be devoted to source subtraction depends critically on the
source counts at the flux levels being probed. There are no good deep
source counts available at 15~GHz; however there exist $\mu$Jy source
counts at 8.4 GHz, less than a factor of two in frequency away from
where AMI would operate, allowing us to assess roughly the level of
source confusion. Windhorst et al \shortcite{windhorst} find the
8.4~GHz source counts below 1 mJy are described by
$$
N(> S) = (3.57 \pm .57)\; (S_{8.4}/{\rm 1 \, Jy})^{-1.3\pm 0.2}\;\rm sr^{-1}.
$$ 
Using the effective spectral index of $\alpha = 0.53$ (taking into
account the dispersion in their measured spectral indices),
extrapolation to 15 GHz gives
$$
N(> S) = 2.6\; (S_{15}/{\rm 1 \, Jy})^{-1.3}\;\rm sr^{-1}.
$$ 

To calculate the effect of source confusion on AMI, we split the
baselines into two sets. The first is dedicated to high temperature
sensitivity mapping of arc-minute scale CMB structure, and so
comprises only baselines shorter than $2~\rm k\lambda$; these are all
45 baselines from the 3.7-m antennas and 6 baselines between the 13-m
antennas.  This set has a flux sensitivity of $51~\mu \rm
Jy~(12~hours)^{-1/2}$ over a 0.1 square degree field.  The second
measures the flux densities of confusing radio sources and comprises
the remaining 22~baselines from the 13-m antennas. It has a flux
sensitivity of $35~\mu \rm Jy$ over the same area in the same
time. The residual confusion noise in the maps made with the compact
set is given by
$$
\sigma_{\rm conf}^2 = \Omega \int_0^{S_{\rm sub}} S^2 \frac{{\rm
d}N}{{\rm d}S}{\rm d}S,
$$
where $\Omega$ is the synthesised beam area and $S_{\rm sub}$ is the
flux level subtracted down to. Assuming a source detection threshold
($4.0\sigma$) of $140~\mu$Jy, we find that the residual confusion
noise in a 12-hour, 1.5-arcmin resolution AMI observation is
30$\mu$Jy, which when added in quadrature to the thermal noise results
in a less than 20\% increase in overall noise level. In practice,
simultaneous fitting of the clusters and point sources to all the
baselines will increase the effective sensitivity to sources and allow
them to be subtracted to an even lower level.

This design achieves optimal sensitivity to the cluster SZ effect and
a separation with other components such as radio sources.  Although a
main aim is the study of clusters, it will probe generally the
structure of the CMB on scales smaller than those accessible to the
Planck satellite. It is sensitive to phenomena such as inhomogeneous
ionisation, density--velocity correlations (Ostriker-Vishniac effect),
filaments and topological defects, which are all of immense interest
as well. In the following sections, however, we demonstrate the
ability of AMI to discover clusters.

\section{Simulated SZ Cluster Sky Maps}
\label{sec:model}

\begin{figure*}
\epsfig{figure=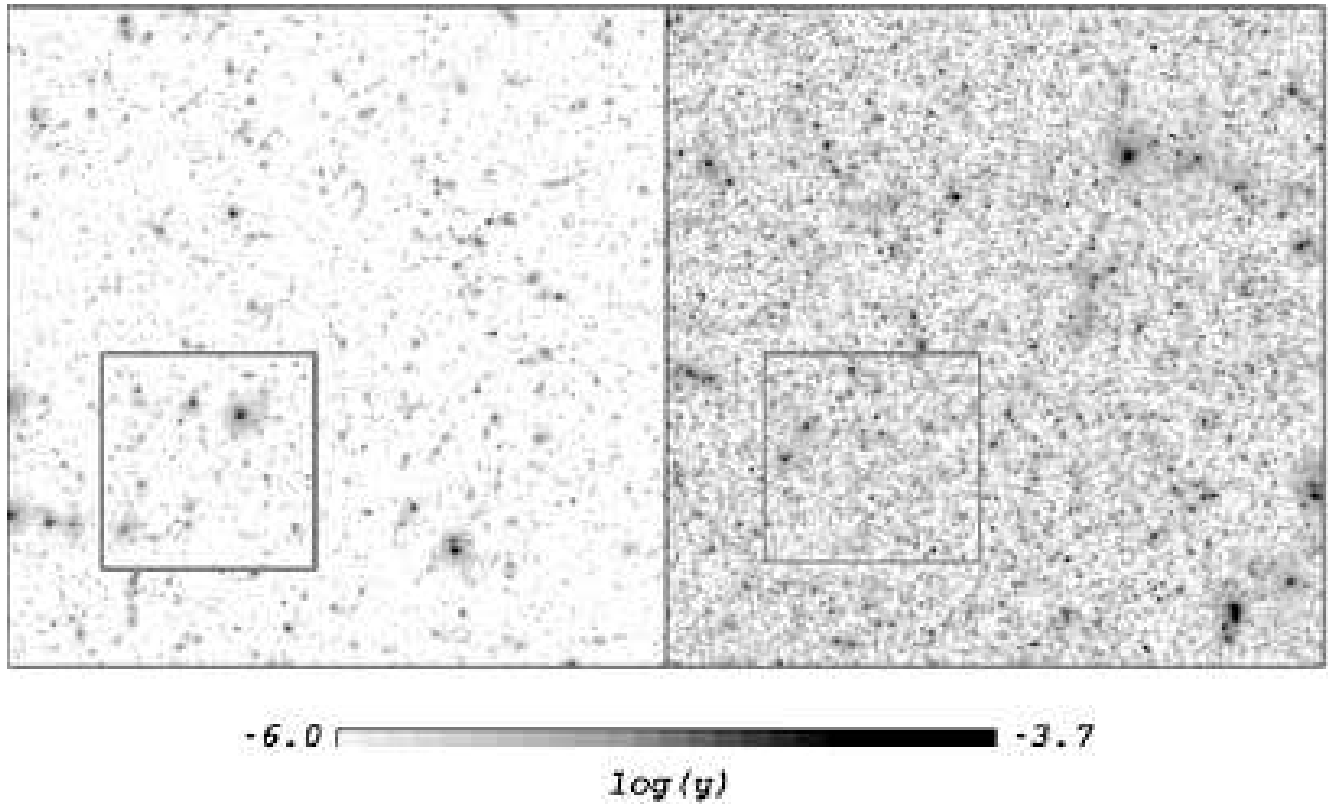,width=11cm,angle=-90}
\caption{\small SZ cluster maps (0.75 arcmin pixels, $5\deg \times 5\deg$), 
simulated as described in the text, with $\Omega_{0} = 1$ (left) and 
$\Omega_{0} = 0.3$ (right). AMI observations of the framed fields 
have been simulated in detail and are presented in Figure~\ref{fig:survey}.}
\label{fig:szmap}
\end{figure*}

To determine how many galaxy clusters AMI might detect, 
it is useful to create realistic SZ sky maps 
simulated for different cosmologies. This approach is 
particularly helpful for interferometric measurements for the 
following two reasons. 
Firstly, limited coverage in the aperture (uv) plane produces ringing in 
real space, where clusters are most easily identified. Thus, the resulting
point-source response in the interferometer observation, the
synthesised beam, has a complicated wing structure that 
produces correlated noise. The extent 
to which this impacts upon the cluster detection efficiency is
difficult to gauge without simulations of the instrumental response 
applied to realistic input sky maps.
Secondly, because interferometers are sensitive to the detailed shapes
of the structures they observe, it is important that the simulated
clusters should have a realistic distribution of shapes and sizes.

Two extreme cases of cluster evolution are considered here by assuming 
a flat and an open 
universe with $\Omega_{0} = 1$ and $0.3$ respectively. The inclusion of a
$\Lambda$--like term does not greatly change the cluster number counts
(eg. Eke, Cole \& Frenk 1996), so this complication is not introduced into the
models.  A Hubble constant of 70 km s$^{-1}$ Mpc$^{-1}$ is chosen for both
cases, and the assumed cold dark matter power spectrum of density fluctuations
is described by a shape $\Gamma=0.25$ (Bardeen et al. 1986) and a linear
theory rms amplitude of mass fluctuations matching the present-day cluster
abundance as given by
\begin{equation} 
\sigma_8 = (0.52 \pm 0.04) \, \Omega_0^{-0.46 + 0.10 \, \Omega_0}, 
\end{equation} 
(Eke et al. 1996).

\begin{figure*}
\epsfig{file=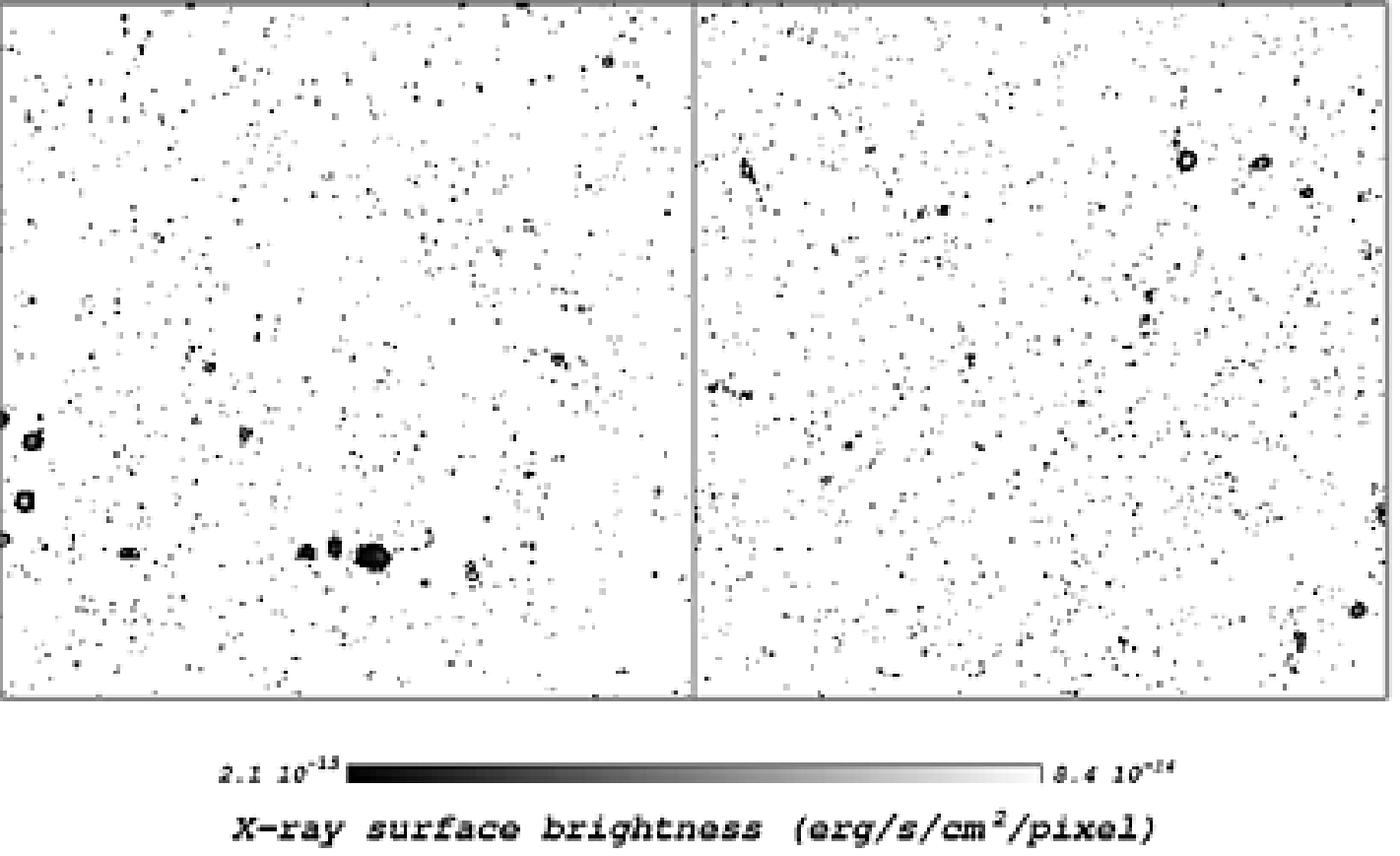,width=11cm,clip=,angle=-90}
\caption{The 0.5--2 keV X-ray corresponding to Figure~\ref{fig:szmap}. 
The X-ray surface brightness limits of the greyscale bar 
($2.1 \times 10^{-15}$;$8.4 \times 10^{-14}$) correspond to 
0.1 and $2.5 \times 10^{-3}$ XMM-Newton (EPIC) counts s$^{-1}$ and per 
(0.56 arcmin$^{2}$) pixel. The lower limit corresponds to the sensitivity 
of a medium-deep XMM survey. Note that the grayscale is inverted against the 
background to enhance the contrast. Only a minimal difference between the 
cosmologies ($\Omega_{0}=1$ to the left and $\Omega_{0}=0.3$ to the right) 
is visible even at this low surface brightness limit, 
because the redshift selection (see Figure~\ref{fig:redcomp}) 
is so different to the SZ effect.} 
\label{fig:xraymap}
\end{figure*}

The above parameters define the background cosmological models and the dark
matter properties. In order to create SZ sky maps, it is also necessary to
specify the fraction $f_g$ of mass that is in the intracluster gas, and the
conversion between total cluster mass and gas temperature as parametrised by
$\beta$ in
\begin{eqnarray}
\lefteqn{\frac{k T_{gas}}{\rm keV} =} \nonumber \\
 & & \frac{7.75}{\beta} 
\left( \frac{M}{10^{15} h^{-1} {\rm M}_\odot}\right)^\frac{2}{3} 
\left( \frac{\Omega_0}{\Omega(z)} \right)^\frac{1}{3} 
\left( \frac{\Delta_c}{178} \right)^\frac{1}{3}(1+z). 
\label{eq:mt}
\end{eqnarray}
$M$ represents the cluster virial mass and $\Delta_c$ is the ratio of mean
halo density to critical density at the redshift of observation, $z$in a
spherical collapse model. As the electron density in the intracluster gas is
proportional to $f_g$, and the cluster SZ decrement is proportional to the
integrated line-of-sight electron pressure, the choice of $f_g/\beta$ has a
significant impact upon the anticipated signals from the clusters. While the
value of $\beta$ can be estimated as $\beta \approx 1$, the value adopted
here, to an accuracy of the order of 10 \% from hydrodynamical simulations
(Bryan \& Norman 1998), there is a somewhat larger uncertainty in $f_g$.  If
one chooses to set $f_g$ using the measured intracluster gas fractions, then
the value should lie in the range given by Ettori and Fabian (1999) and Mohr,
Mathiesen and Evrard (1999), who both find 0.1--0.25 at the 95 \% confidence
level.  An alternative way to choose $f_g$ would be through the primordial
nucleosynthesis value for the fraction of the critical density contributed by
baryons, $\Omega_b \approx$ 0.019 $h^{-2} \approx 0.04$ (Tytler et
al. 2000). For the two different $\Omega_0$ values, the assumptions that
clusters contain the universal baryon fraction (White et al. 1993) and that
all baryons are in the gaseous component lead to $f_g=0.04$ and $0.13$.  A
value of $f_g=0.1$ has been used for both maps described in this paper.  This
is on the low side unless $\Omega$ equals 1, and primordial nucleosynthesis is
a more appropriate way to determine $f_g$ than studying clusters
themselves. Thus, the predicted cluster 
number counts are likely to be lower than 
might otherwise be expected. Furthermore, the choice of the same $f_g$ for
both cosmologies will reduce the difference between the total numbers of
detectable sources relative to what would be found with $f_g$ inferred from
nucleosynthesis.  The model parameters are summarized in 
Table~\ref{table:params}. 

To produce a map, the Press-Schechter expression (1974) is used to create a
list of cluster masses and redshifts. The centres of these clusters are placed
at random within a 5$^\circ$ $\times$ 5$^\circ$ sky map with 45 arcsec pixels,
and template cluster maps are pasted, suitably scaled, onto these
positions. Individual cluster templates are produced from the ten $\Lambda$CDM
hydrodynamical cluster simulations of Eke, Navarro and Frenk (1998). Each
cluster is observed at eight different redshifts out to $z\approx 1.1$, and
templates are produced from three orthogonal directions to maximise the
variety in the apparent cluster shapes.  At $z=0$ these clusters have total
masses of $\sim 10^{15} M_\odot$ and thus represent the largest virialised
structures.  If a smaller cluster is required then the templates are scaled
down in angular size, gas temperature and SZ emission accordingly. The
discrete redshift sampling of the simulations means that some scaling of the
template emission to the desired redshift is also necessary. The two simulated
SZ maps are shown in Figure \ref{fig:szmap}.

\begin{table}

\begin{center}
\begin{tabular}{c|cc}
$\Omega_{0}$ & 1.0 & 0.3 \\
$h$ & 0.7 & 0.7 \\
$\sigma_8$ & 0.52 & 0.87 \\
$\Gamma$ & 0.25 & 0.25 \\
$f_g$ & 0.1 & 0.1 \\
$\beta$ & 1 & 1 \\
$X$ & 0.76 & 0.76 \\
($\Omega_{\rm b}h^2$ & .049 & .0147) \\
\end{tabular}
\caption{Summary of model parameters. The 
universal gas fraction is $f_g = m_b / m_{DM}$, 
$\beta = \mu m \sigma^2 / kT$ is the ratio of 
kinetic to thermal energy in the cluster gas, and $X = m_{\rm H} / 
m_{\rm total}$ the hydrogen~/~helium ratio.}
\label{table:params}
\end{center}
\end{table}

da Silva et al. (2000) simulated SZ sky maps by stacking together
hydrodynamical simulation boxes, in order to gain sufficient depth in
redshift. Once the different choices of the various parameters that are
described above are taken into account, $\sigma_8$ being of particular
importance, their mean flux decrement per pixel is $\approx 30$\% greater than
that found in the maps produced here.  Decreasing the minimum SZ flux of the
included clusters from $Y_{min} = 4 \times 10^{-7} h$ arcmin$^{2}$ largely
accounts for the difference in mean flux. $Y$ is the integral of the Compton
$y$ parameter over the cluster solid angle (ie area divided by the angular
diameter distance squared, $Y=r_d^{-2} \int y \, {\rm d}A$). However, this
difference is less than 10 \% of the flux of the faintest detected clusters,
so this does not significantly influence the number of clusters that are
detected here.

X-ray cluster emission templates were also produced using the same
simulated hydrodynamical clusters as for the SZ templates. Bolometric
luminosities were calculated from the particle densities and
temperatures according to equation (15) of Eke et al. (1998). This
allowed the corresponding X-ray maps to be created, so that the same
simulated skies could be observed at other wavelengths. 
0.5--2~keV and 2-10~keV maps were created, with the cluster 
luminosities being scaled from the template temperature to that of the 
required cluster assuming the non-evolving 
\begin{equation}
L_x \propto T^3
\label{eq:lxt}
\end{equation}
(Ettori, Allen \& Fabian 2001, Donahue et al. 1999, Della Ceca et al. 2000, 
Schindler 1999, Fabian et al. 2001), rather than the 
$L_x \propto T^2 (1+z)^{3/2}$ that is expected from simple scaling arguments. 
Possible evolution affects our modelling only very weakly, 
since we rescale within a redshift bin. 
The 0.5-2 keV X-ray cluster map is complete to an X-ray flux limit of 
$1 \times 10^{-15}$ erg cm$^{-2}$ s$^{-1}$. 

\section{Detecting clusters with AMI}
\label{sec:clusters}

We simulate the cluster detection process for a blank field AMI
observation in two different ways: a detailed, but time consuming,
simulation of the response of the interferometric array to 
the structure on the sky, and a simplified simulation,
which only operates in the image plane and uses a compensated beam profile 
that is constructed to match the synthesized beam resulting from the more
detailed simulations. 

\begin{figure}
\centerline{\epsfig{file=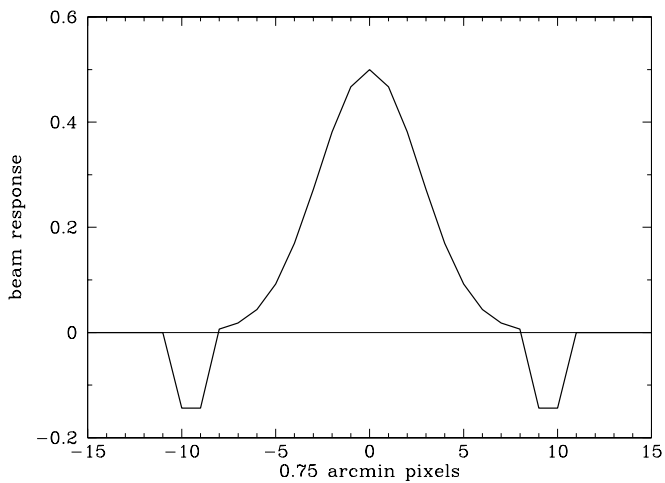,height=7cm}}
\caption{A model for the synthesised beam of AMI. The beam efficiency, 
central height, is 0.5, FWHM is 4.8 arcmin and the only other two 
parameters, since the beam is exactly compensated, the inner and outer 
radii of the ring are adjusted to be at 3 and 4~$\sigma$ of the Gaussian 
(6.12 and 8.16 arcmin) respectively. When this beam model is convolved 
with a typical cluster profile it gives a good approximation to the flux 
observed in the detailed simulations (Figure~7).}
\label{fig:beamm}
\end{figure}

\begin{figure}
\centerline{\epsfig{file=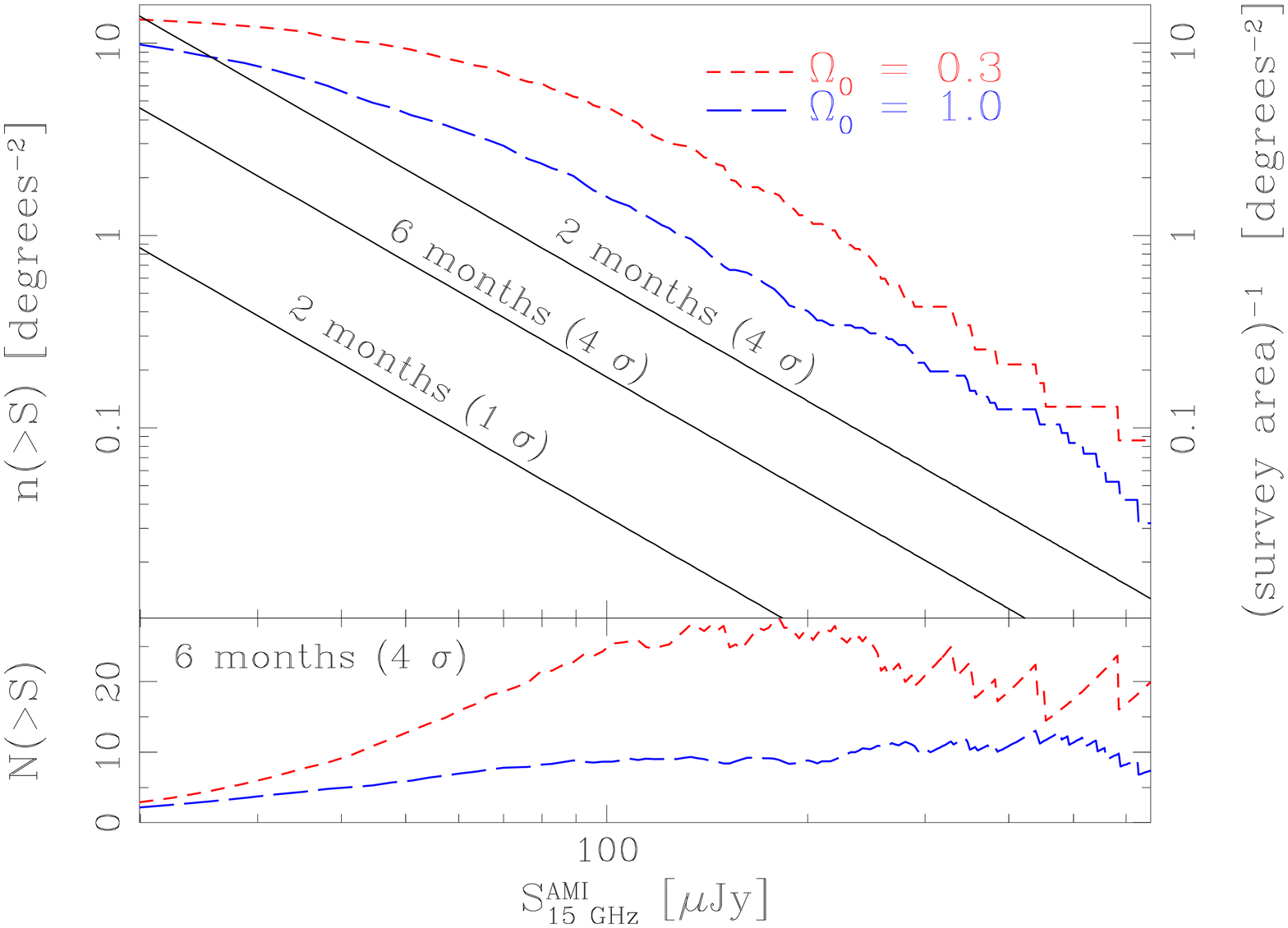,width=8.5cm}}
\caption{Upper panel: Observed AMI SZ cluster cluster source counts for low 
(short dashed; top) and high (long dashed; below) matter density and 
survey sensitivity lines (solid; labeled) in (sky area)$^{-1}$ as a function 
of observation time. The ratio between these curves equals the number of 
detected clusters (lower panel). At high fluxes the cluster counts rise 
steeper than the sensitivity lines, showing that a deeper survey will 
return more clusters for the same observation time up to the point where 
confusion sets in. Thus a broad maximum in the expected number of 
clusters is seen at around a few hundred $\mu$Jy for both cosmologies. 
However the cluster counts can also be tested with good efficiency 
over a wide range of fluxes by surveying differing areas at appropriate 
depths. Note that a shift in the position 
of the maximum to lower fluxes with decreasing matter density is in fact 
expected, since the increasing number of high redshift clusters steepens 
the cluster counts which are normalised at high fluxes 
(cf. Figure~\ref{fig:paradep}). 
The error on the ratio of the curves and the field-to-field variation is 
approximated well by Poisson statistics for cluster numbers.} 
\label{fig:ncounts}
\end{figure}

\begin{figure*}
\mbox{
  \epsfig{figure=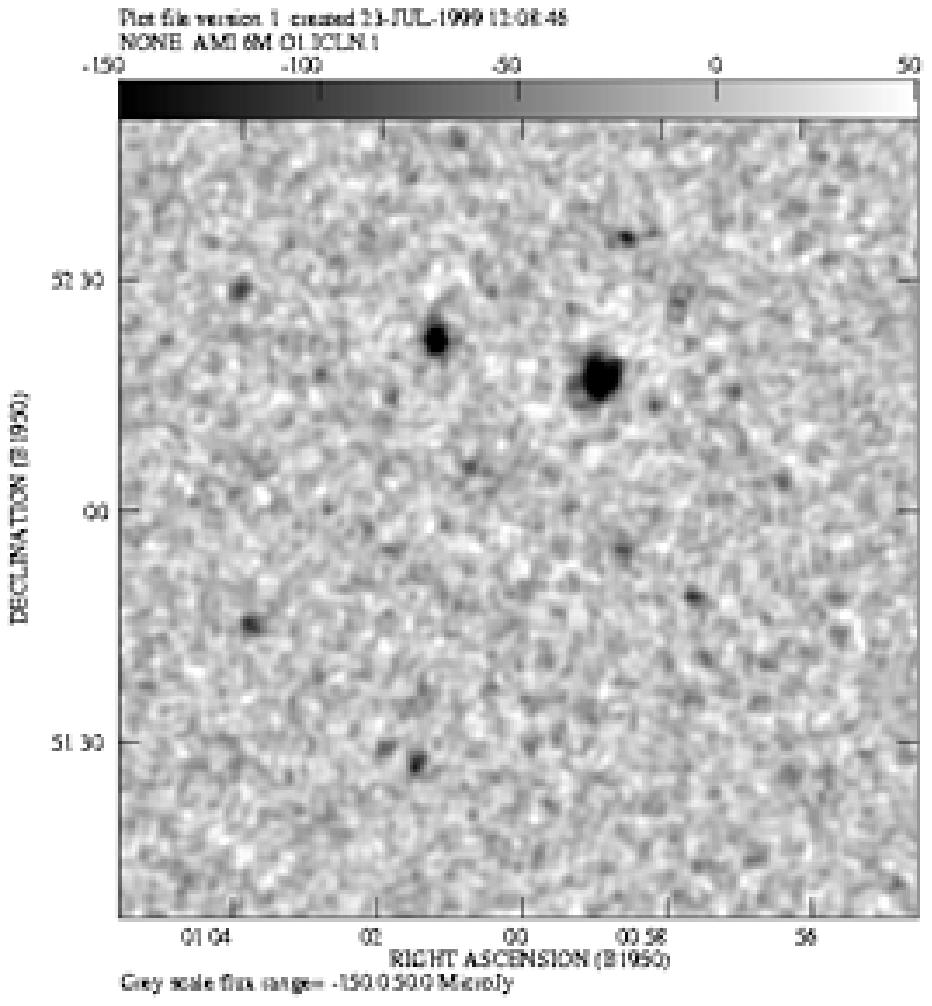,width=9cm,angle=0,%
bbllx=0pt,bblly=10pt,bburx=268pt,bbury=287pt,clip=}
  \hspace{0cm}
  \epsfig{figure=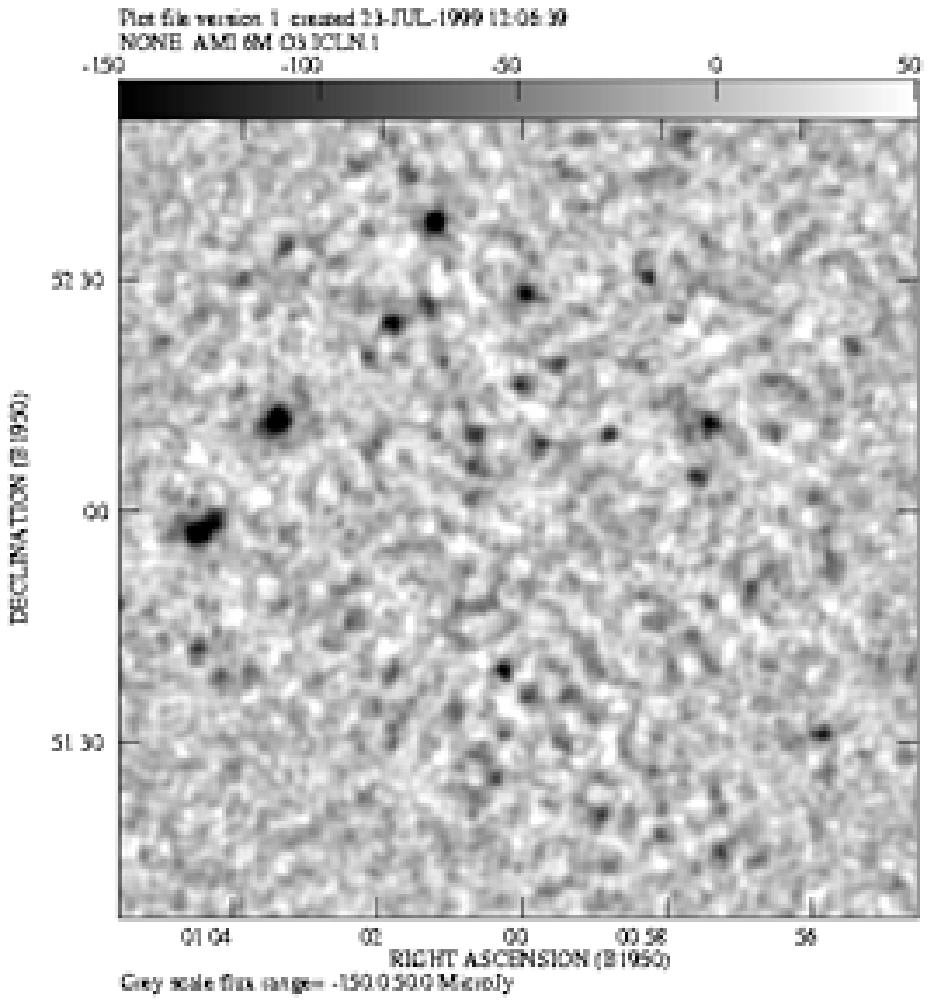,width=9cm,angle=0,%
bbllx=0pt,bblly=10pt,bburx=268pt,bbury=287pt,clip=}
}
\caption{Two simulated AMI observations of 20-arcmin radius fields 
showing the effects of the different cosmologies ($\Omega_0=1$ on the 
left and $\Omega_0=0.3$ on the right; units are $\mu$Jy beam$^{-1}$. 
The clusters show as dark (negative) features against the CMB. 
The simulation with low matter density has many more moderate mass 
clusters, because clusters form early and we can see them in SZ 
all the way to very high redshift.}
\label{fig:survey}
\end{figure*}

The simplified procedure, aims at determining the cluster 
counts resulting from mosaiced observations, and with a realistic beam, 
in a fast and straightforward way. The synthesised 
beam depends on the coverage in the visibility plane which itself depends 
on the array configuration, the sky position of the field and 
the length of observations. The finiteness of the coverage causes 
ringing in the Fourier transform and the wing-structure of the beam. 
An accurate analytic description can be given with Bessel functions, 
which however still requires knowledge of the visibility coverage and 
substantial computational efforts. However a simpler model for 
the beam with only a few parameters can be given by a Gaussian 
beam with a compensating negative ring (see Figure~\ref{fig:beamm}), 
which approximates the most relevant parts of the synthesised beam 
surprisingly well. Convolving the simulated SZ cluster map with this 
beam results in a map in which the brighter pixels give the positions 
and fluxes of the brightest clusters; others can be confused. 

After convolution of the cluster sky maps with this beam the observed
cluster fluxes and the integrated cluster counts can be 
constructed (see Figure~\ref{fig:ncounts}). Here a beam width of 4.5
arcmin FWHM and a beam efficiency of 0.5 were assumed. In addition to
the recovered cluster counts we show the 1-$\sigma$ and 4-$\sigma$
sensitivity lines for various observation times. We assume
a Gaussian thermal noise floor corresponding to a system temperature
of 30~K.

\begin{figure}
\centerline{\epsfig{file=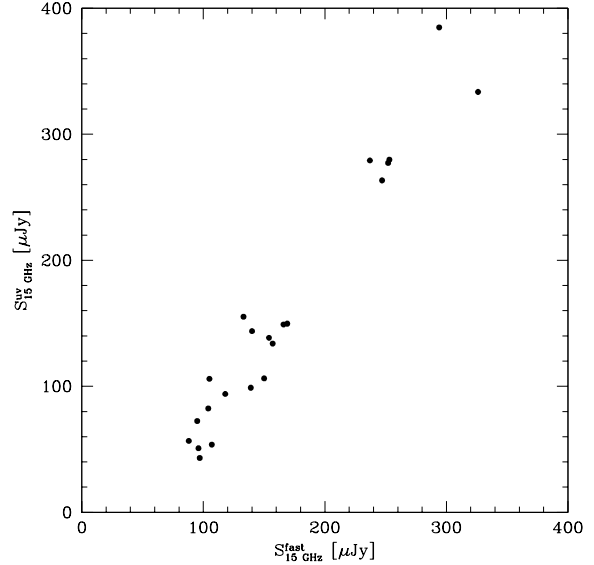,height=8cm}}
\caption{Comparison between the two algorithms simulating AMI observations; 
the fast algorithm on the horizontal axis against the detailed simulation 
of a small field on the vertical axis. Both methods estimate the recovered 
cluster flux sufficiently accurately for our purpose of evaluating the 
prospects of cluster detection, and we are not addressing here the issue 
of the final flux accuracy in the survey.}
\label{fig:fluxcomp}
\end{figure}

The detailed telescope model takes all baselines of a given array
configuration, and calculates the uv coverage for an observing run.  The
simulated sky images are multiplied by the interferometer primary beam,
Fourier transformed, and then sampled at the calculated positions in the uv
plane. Gaussian noise is added to the data points at the appropriate
level. The resulting `observed' data are then used to produce a map (see
Figure~\ref{fig:survey} for an example) in the same way real observations are
analysed, i.e. with {\sc Aips} tasks, and the detected clusters are extracted
from the CLEANed map. We compare the cluster fluxes so derived with those from
the fast procedure (see Figure~\ref{fig:fluxcomp}) and find variations and
scatter between the two methods that are small in relation to the
uncertainties in the cluster sky model parameters.  We conclude that the
predicted cluster counts are not affected by these differences.

\section{Determination of Model Parameters}
\label{sec:parameters}

\subsection{Based on AMI cluster counts}

\begin{figure}
\centerline{\epsfig{file=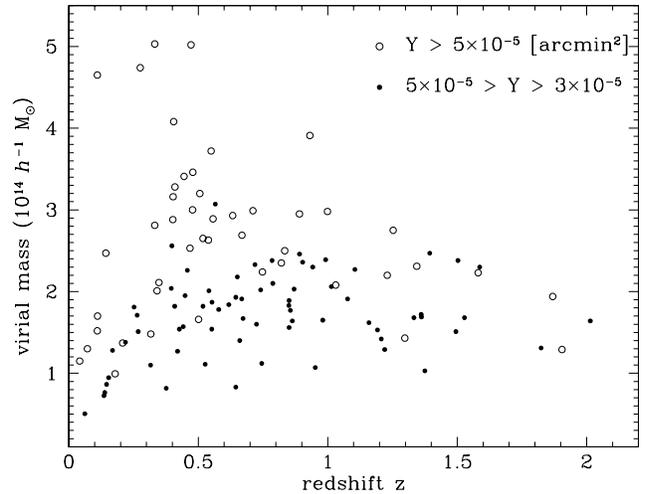,width=7cm,angle=-90}}
\caption{Masses and redshifts of the clusters detectable in the AMI survey in
the $\Omega = 0.3$ cosmology.  The AMI flux limit corresponds to an 
almost constant mass limit over all redshifts. The (arbitrary) flux binning
demonstrates the advantage of a deep survey in finding high redshift
clusters.}
\label{fig:mzdist}
\end{figure}

\begin{figure}
\centerline{\epsfig{file=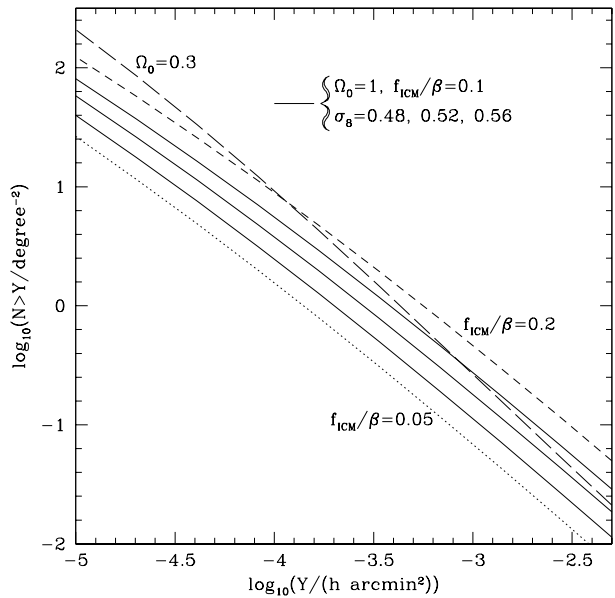,width=8.5cm,angle=0}}
\caption{Cumulative source counts above a given integrated SZ effect, $Y$.
The three solid lines illustrate the effect of varying the power spectrum
normalisation, short-dashed and dotted lines show how the counts vary with
increasing and decreasing gas fraction, and the long-dashed line is the
default $\sigma_8$ and $f_{ICM}/\beta$ prediction for the $\Omega_0=0.3$
model.}
\label{fig:paradep}
\end{figure}

After one year of observation AMI will have completed an 
unbiased cluster survey down to roughly $10^{14}$ M$_\odot$ 
with 20 ($\Omega_{0}$ = 1) to 70 ($\Omega_{0}$ = 0.3) 
clusters for our pessimistic assumptions about $\sigma_8$ and 
$f_g$, or several hundred clusters for more realistic assumptions. 
The SZ cluster catalogue and the differential cluster counts at 
different flux levels will be prime observational results 
from the AMI blank field survey. Figure~\ref{fig:mzdist} shows the 
mass--redshift distribution we expect for the catalogued clusters 
in a $\Omega_{0}=0.3$ cosmology. The plot demonstrates that the 
sample is nearly mass-limited and that in a deep survey, which AMI 
can provide, many high redshift clusters can be found. 
This observational result, the cluster catalogue, needs to be interpreted 
in the framework of a model of cluster evolution. 
In our model, degeneracies exist between the input parameters 
for the observed cluster counts. 
Assuming prior errors on the input parameters as given in 
Table~\ref{tab:paradep}, a distinction between the two cosmologies 
or a reliable determination of $\Omega_{0}$ is only marginally 
possible. Note however that the uncertainty in $\sigma_8$ comes mainly 
from the uncertainty in the mass--temperture conversion when inferring 
$\sigma_8$ from the cluster temperature function. For our purposes 
in predicting a mass $\times$ temperature function the uncertainty 
given in the Table can be seen as an upper limit. Extreme values for 
the individual parameters will already be severely constrained 
by the observation. At AMI sensitivities of 
$r_d^{-2} \int y \, {\rm d}A = Y \approx 4 \times 10^{-5}$ arcmin$^2$, the 
expected change in cluster counts due to changing the matter density 
is $N(\Omega_{0} = 0.3) / N(\Omega_{0} = 1) \approx 3.5$. 
In Figure~\ref{fig:paradep} we show the parameter degeneracy as a 
function of the flux limit, and note that the matter density is 
better determined by cluster counts the deeper the observation is, because 
the slope of the counts is a function of the matter density as well. The 
Planck Surveyor, for example, is complementary to AMI in probing the counts 
at a higher flux limit of about 3--7~$ \times \, 10^{-4}$ arcmin$^2$, where 
the cluster counts are mainly determined by $\sigma_8$, and the gas fraction. 
The slope of the cluster counts measured over a wide range and to low fluxes 
would give direct evidence of the value of $\Omega_{0}$. 

\begin{table}
\begin{center}
\begin{tabular}{|lll|}
Parameter & change in & fractional change \\
 & percent & in $N(>Y)$ \\
\hline
$h$ & 20 \% & 1.3 \\
$f_g$ & 30 \% & 1.5 \\
$\sigma_8$ & 7 \% ($\sim$ 1 $\sigma$)& 1.5 \\
 & 14 \% ($\sim$ 2 $\sigma$) & 3.2 \\
\end{tabular}
\end{center}
\caption{The change in cluster counts by varying the model parameters.}
\label{tab:paradep}
\end{table}

In practice however, even with deep surveys the confusion between parameters
remains unsatisfactory, but can be broken easily with basic follow-up
observations in optical and X-ray wavebands, as we detail in the next section.

\subsection{Basic X-ray and optical follow-up}
\label{sec:followup}

\begin{figure}
\centerline{\epsfig{file=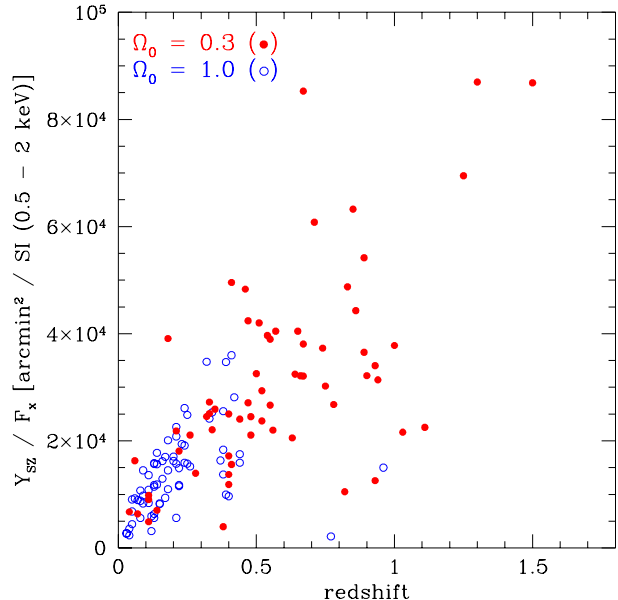,width=8.5cm,angle=0}}
\caption{The ratio between SZ and X-ray flux (taken with a large aperture 
comparable to the virial radius for most clusters) is redshift dependent, 
``photometric'' redshift; details in the text.}
\label{fig:szxray}
\end{figure}

\begin{figure}
\centerline{\epsfig{file=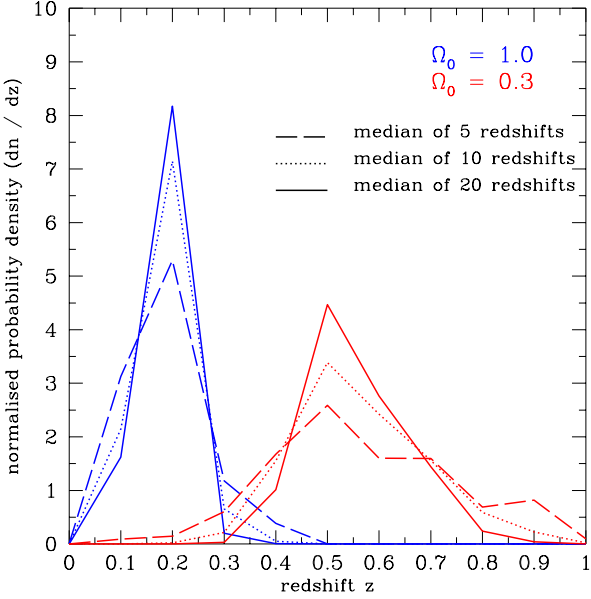,width=8.5cm,angle=0}}
\caption{The distributions of median redshifts for the clusters detected 
by AMI separate the two cosmologies better with increasing sample size. 
However when redshifts are measured for only half of the clusters of a 
sample of only 20 clusters and lower limits exist for the other clusters, 
then there is a 90~\% chance that one of the two cosmologies can be ruled 
out with $>$99.9~\% confidence.} 
\label{fig:redshift}
\end{figure}

\begin{figure}
\centerline{\epsfig{file=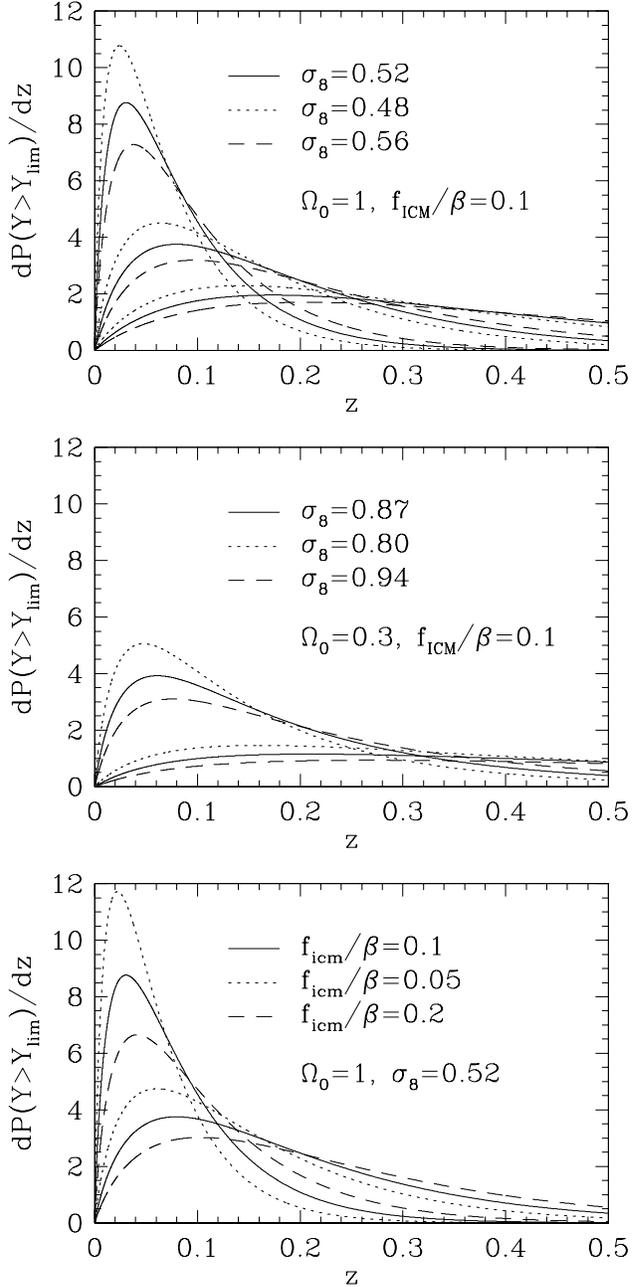,width=8.5cm,angle=0,
bbllx=130pt,bblly=95pt,bburx=440pt,bbury=724pt,clip=}}
\caption{Redshift distributions for the model cluster distributions assuming
various different model parameters. The top panel shows the effect of
varying the power spectrum normalisation $\sigma_8$ for the $\Omega_0=1$
model with the default value of $f_{ICM}/\beta$. The three different sets
of curves correspond to different limiting cluster Y values of $10^{-3},
10^{-4}$ and $10^{-5}$ h arcmin$^2$ going from most peaked to most
extended redshift distribution respectively. The middle panel shows the
corresponding figure for the $\Omega_0=0.3$ model. Only the
$Y_{lim}=10^{-3}$ and $10^{-4}$ cases are shown, with the fainter clusters
being even more extended. The bottom panel shows how the redshift
distribution changes with $f_{ICM}/\beta$ for the cluster normalised
$\Omega_0=1$ model. Again, only the $Y_{lim}=10^{-3}$ and $10^{-4}$ cases
are plotted.}
\label{fig:parared}
\end{figure}

Measuring redshifts optically and using the
X-ray to $Y$ flux ratios to identify the high redshift clusters
(cf. Figure~\ref{fig:szxray}) will allow an estimate of the cluster space
density as a function of the thermal energy, which is 
almost the mass, and of the 
redshift. The redshift distribution, crudely speaking, depends on $\Omega_{0}$
and cluster physics affecting the gas fraction and the temperature both as a
function of redshift.  In Figure~\ref{fig:redshift} we demonstrate the
discriminative power of a statistical measure based on the redshift
distribution, in this case the median, to distinguish values of $\Omega_{0}$
even for a small cluster sample and incomplete follow-up, as long as redshift
limits exist for half of the sample. The statistical analysis can be refined
with better data to the point where a direct fit to the redshift data is
possible, and tests of the very assumptions underlying our model can be
carried out, i.e. for Gaussian initial conditions, gravitational collapse and
various cluster physics effects neglected in the hydro-simulations. In
Figure~\ref{fig:parared} we show that the redshift distribution is a much more
robust estimator for the $\Omega_{0}$ than the cluster counts, affected only
by variations in the other model parameters larger than their presumed
uncertainties.

\begin{figure}
\epsfig{file=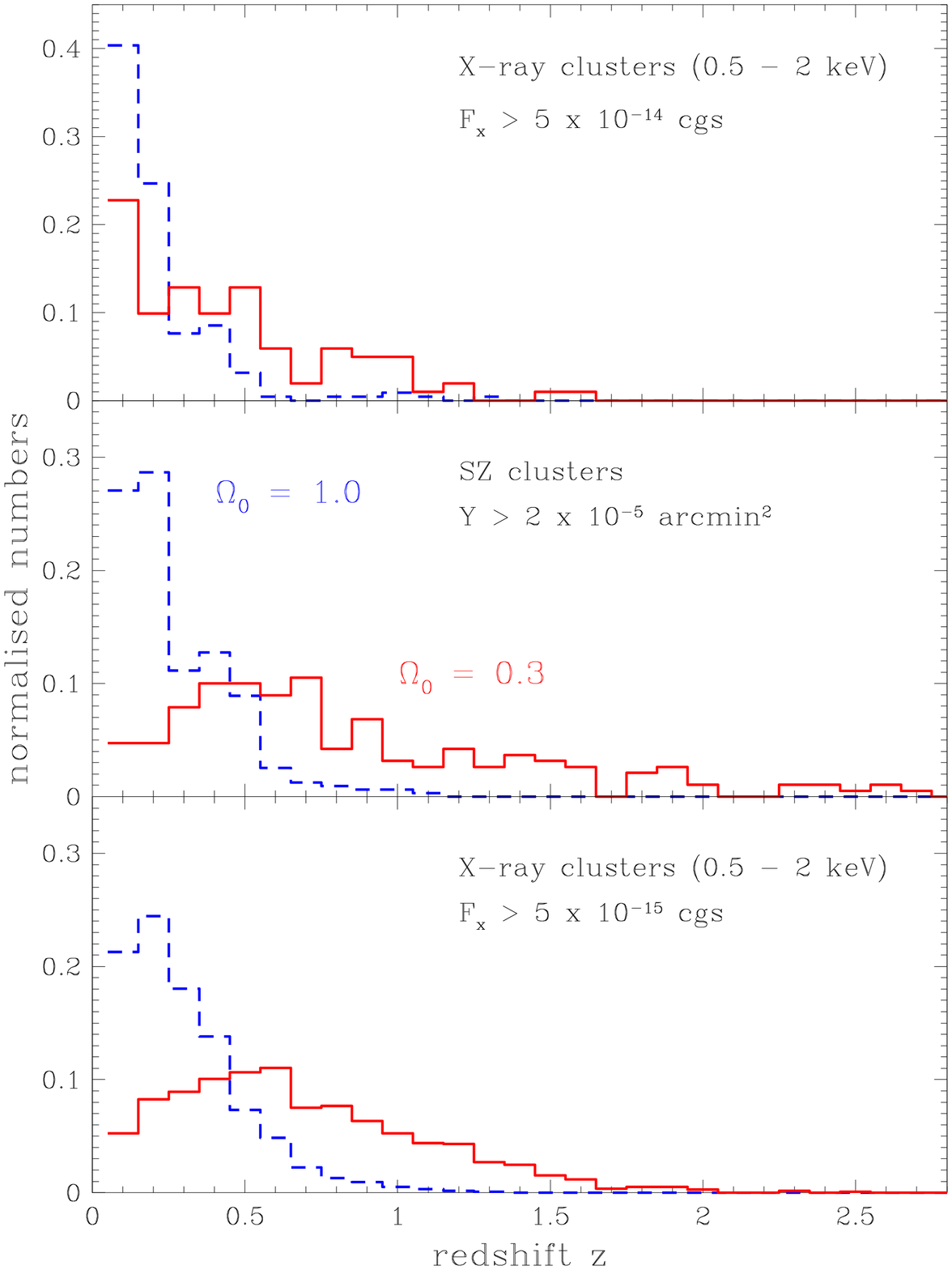,width=8.5cm,clip=}
\caption{Comparison of the redshift distributions for cluster samples selected
from SZ and X-ray observations. The depth in X-ray flux in the top panel is a
factor of two below current flux limits in medium sized surveys and returns of
the order of 4 clusters per square degree, a number similar to the SZ case
(centre panel).  However to distinguish the cosmologies comparable to the SZ
case another factor of 10 deeper (bottom panel) is required. Even at this low
flux limit the X-ray selection process returns relatively fewer clusters at
very high redshifts than the SZ observation. Note the near completeness in all
samples for the $\Omega_{0} = 1$ case.}
\label{fig:redcomp}
\end{figure}

With X-ray observations in particular we will get better determined cluster
positions, the X-ray structure, indications of the gas temperature from
hardness ratios and in a few cases redshifts via the iron-K line. The ratio of
$Y$ to X-ray flux (see Figure~\ref{fig:szxray}), directly proportional to the
cooling time of a cluster at a given redshift, is also an indicator of the
cluster redshift, due to the dimming of the X-ray flux, and the accuracy
increases with redshift, in contrast to optical methods.  With the scaling
relations we have used to produce the sky maps (equations \ref{eq:mt} and
\ref{eq:lxt}), we expect a scaling 
\begin{equation}
\frac{Y}{F_x} \propto h \, f_g \, \beta^{-1} \, T^{-1/2} \, (1+z)^{5/2} \, 
\left( \frac{\Omega_0}{\Omega(z)} \right)^{-1/2} \, 
\left( \frac{\Delta_c}{178} \right)^{-1/2}.
\label{eq:szxray}
\end{equation}
The temperature dependence reflects that of the 
cooling function, is weak, and disappears when lowering the 
assumed power index of our non-evolving $L_x-T$ relation from 3 to 2, 
i.e. the value expected from simple scaling arguments. 
The last three factors in the above relation are all of cosmological 
relevance. The separation between the two cosmological models comes 
mainly from the redshift term, due to the different underlying redshift 
distributions, but also the other two terms give small additional 
factors greater than 1 for $\Omega_{0} < 1$. 
Once the redshifts are known for all clusters, the relation can be used 
to determine $f_g \beta^{-1}$ statistically as a function of redshift 
for the sample. 

Our X-ray images (Figure~\ref{fig:xraymap}) show a flux limit 
comparable to the sensitivity for future medium-deep surveys, 
for example with XMM, and no confusing X-ray background has been added. 
The number of clusters detectable in X-rays is large, but the sample 
is strongly biased towards nearby less massive clusters. 
Therefore no obvious difference between the two cosmologies is apparent 
in the X-ray maps. A comparison of the redshift distributions 
is shown in Figure~\ref{fig:redcomp}. The redshift distributions of 
the observed clusters separate clearly with matter density for the 
AMI samples, but do so only at a very low flux limit and only for a very 
large number of clusters in the X-rays. To find high redshift clusters 
efficiently and to probe cosmology, deep SZ surveys have a clear 
advantage. 
However X-ray follow-up on SZ-selected clusters is also important 
to exploit the information which becomes accessible with such 
an SZ survey, for example to separate gas density and temperature. 
The present 
(and future) X-ray telescopes with high resolution, high sensitivity, 
but small fields of view are ideal for this purpose. The exposure time 
per cluster ideally would be matched to ensure a similar number of X-ray 
counts per cluster in the sample to study it in a largely 
redshift-independent way. 

Therefore combining all the information from the blank field SZ
survey, X-ray and optical follow-up promises a determination of all
the model parameters individually, 
assuming that real clusters are similar enough to our simulated ones. 
The AMI sample can be studied further in many ways 
and compared to other probes of cosmology, which would identify 
cluster physical effects which we have neglected and therefore 
improve the understanding of cluster formation and evolution. 

\subsection{Detailed AMI, X-ray, optical and other follow-up} 

\begin{figure*}
\begin{center}
\hbox{
  \epsfig{figure=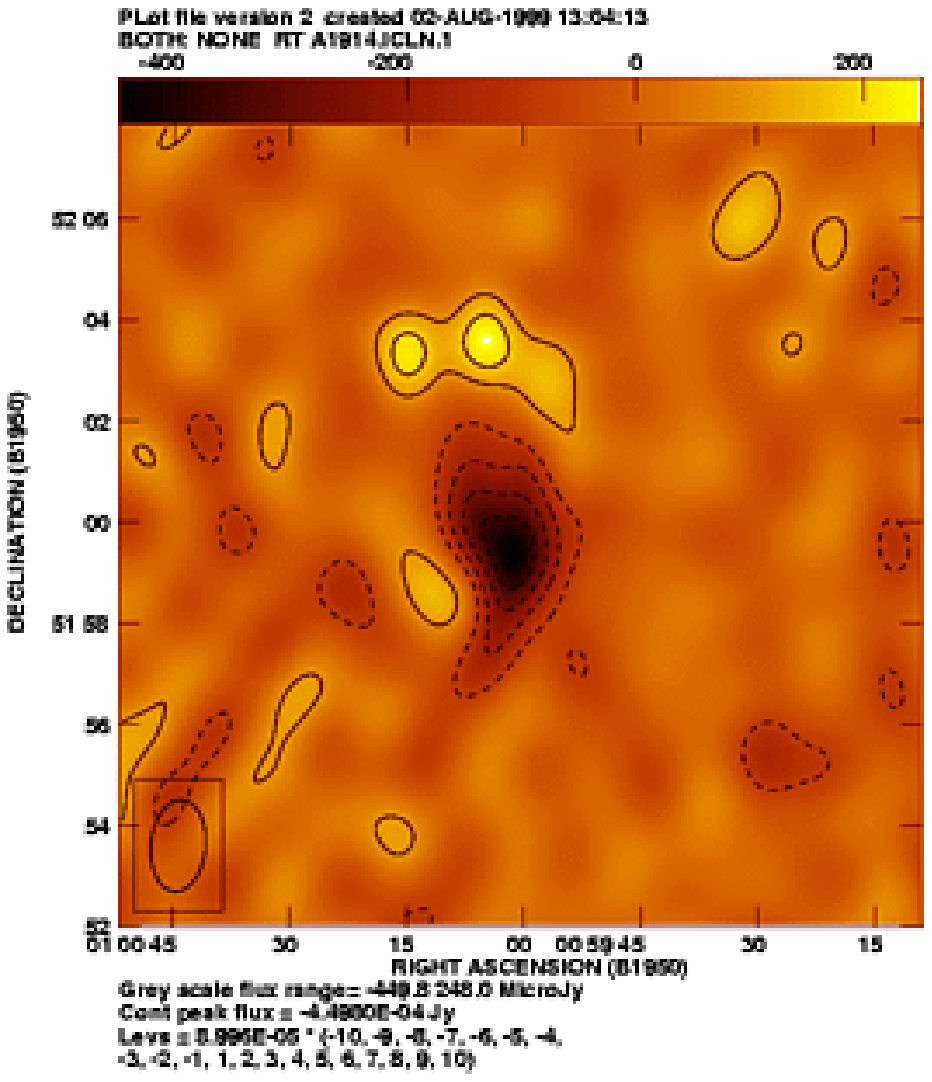,height=9cm,angle=0,%
bbllx=0pt,bblly=17pt,bburx=264pt,bbury=297pt,clip=}
  \hspace{1cm}
  \epsfig{figure=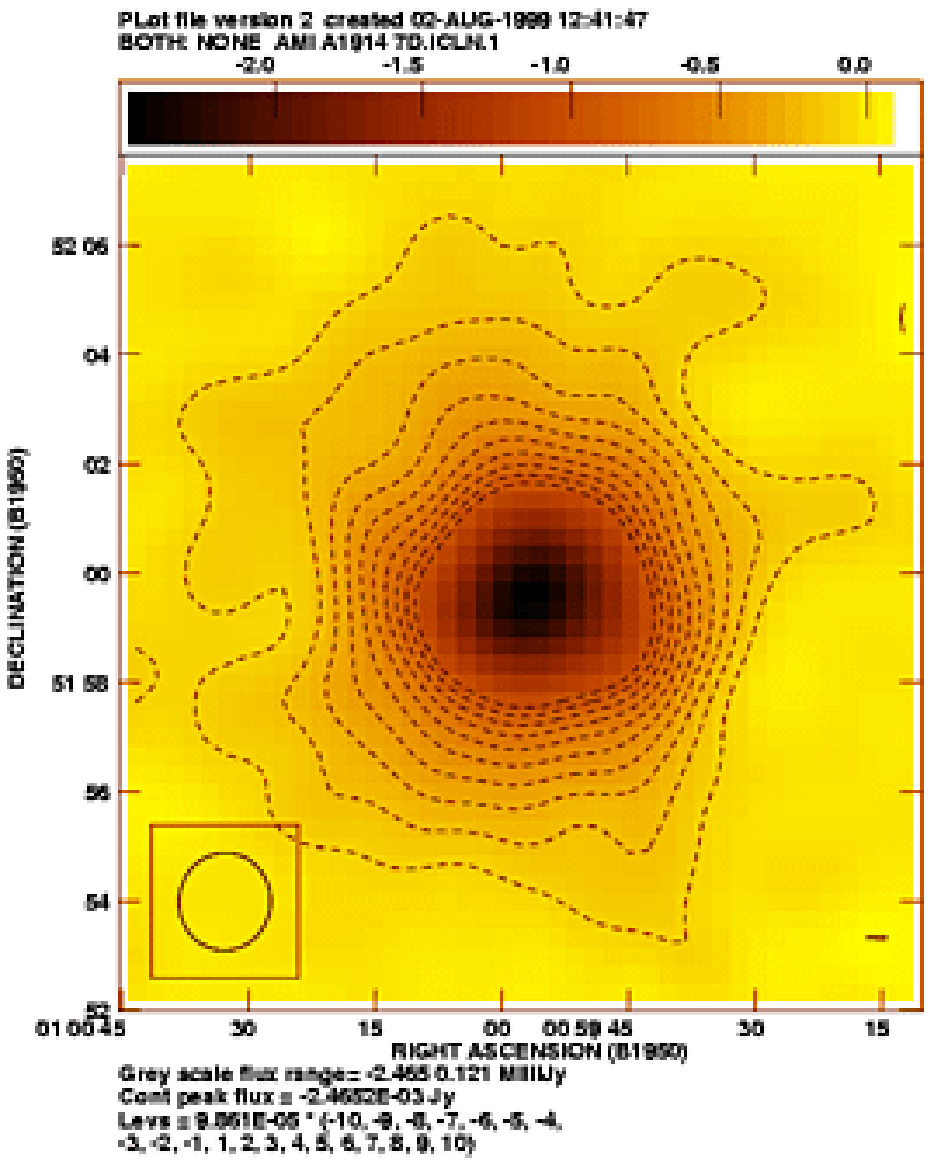,height=9cm,angle=0,%
bbllx=0pt,bblly=17pt,bburx=268pt,bbury=321pt,clip=}
}
\caption{Left: An image of 84 hours of RT data showing an SZ effect in Abell
1914 which is essentially unresolved. Right: A simulated observation of the
cluster with AMI of the same integration time. The spatial dynamic range and
signal-to-noise ratio are vastly increased, providing detailed structural
information on the cluster gas; indeed no structure due to receiver noise is
visible in the background.}
\label{fig:newa1914}
\end{center}
\end{figure*}

The following observations could be carried out to provide a detailed study
of the AMI cluster sample:

\begin{itemize}

\item sensitive, high resolution images from AMI in non-survey mode 
(see Figure~\ref{fig:newa1914}); 

\item X-ray temperature maps; 

\item total projected mass from gravitational lensing; and 

\item SZ observations at frequencies above 217~GHz.

\end{itemize} 

We expect that the following results, all as a function of redshift, 
can be achieved: 

\begin{itemize}

\item gas and temperature structure: basic shapes, 
merging, cooling, structure changes over time; 

\item investigation of the virialisation state 
through SZ and lensing observations; 

\item testing the temperature--mass relation and preheating; 

\item comparing gas and dark matter ($f_g$) structures
(note that the sensitive SZ measurements, which measure the gas mass
directly (as long as the temperature distribution is understood),
would be a better probe of the gas fraction than X-ray measurements,
which give the gas mass in a model-dependent way from the 
emission measure, which is proportional to the square of the density);

\item cluster peculiar velocities (kinematic SZ effect);

\item effects of the (temperature-dependent) relativistic tail 
and non-thermal effects of the electron gas; and 

\item determination of galaxy masses and types.

\end{itemize} 

In particular all the data would be available to determine the angular 
distance $r_d$ to the clusters (Gunn 1978; Silk \& White 1978; Cavaliere, 
Danese and De~Zotti 1979; Birkinshaw 1979), and therefore with the redshifts 
$H(z)$ or to low order $H_0$ and $q_0$ (which depends on $\Omega_{0}$ and 
$\Omega_\Lambda$) can be estimated. 
This absolute distance method is independent 
of cluster evolution and works in principle for an individual cluster, as 
long as the parameters entering the relation for $r_d$ can be measured 
reliably and the cluster gas and temperature distributions can be modelled 
sufficiently accurately. We know there are uncertainties in this 
modelling process, for example through clumping, cooling, or temperature 
gradients. But we have found, from analytic modelling and 
the hydrodynamical simulation templates, that the effects on 
estimating $H_0$ and $q_0$ tend to cancel out even for a single cluster, 
and certainly for a larger sample (Grainger et al.). 
In a low density universe ($\Omega_0=0.3$) where many clusters, 
are expected at high redshifts, 
the AMI sample contains an about equal number of clusters below 
and above $z=0.5$. For simplicity we divide the sample into one 
at $z=0$ which determines $H_0$ and one at $z=0.8$ which in comparison 
determines $q_0$. To achieve an accuracy of about 5 \% in the angular 
distance relation, which is at least comparable to the supernovae results 
(Riess et al. 1998; Perlmutter et al. 1999), we will need samples of 
30 clusters 
each at low and high redshift, mainly to reduce the effect of unknown 
orientation. 
A different and source evolution independent method to constrain the 
cosmological constant seems in sight. 

\section{Conclusions}
\label{sec:conclusions}

Technological advances now allow a detailed study of cluster gas in SZ 
for individual bright clusters and a blank field survey to search for
fainter clusters. Both aspects are of great relevance for
understanding the formation and evolution of clusters and their gas
content in a cosmological context. The SZ effect probes the product of 
gas mass and temperature directly and almost independently of redshift. 

We have shown: 
\begin{enumerate}

\item We have assessed the performance in finding clusters 
of the proposed interferometer array AMI with realistic SZ sky simulations 
and detailed simulations of the observing process. With 
very conservative assumptions about gas fractions and the power spectrum 
amplitude, AMI will 
discover 20 ($\Omega_0=1$) to 70 ($\Omega_0=0.3$) clusters 
with total mass $\ge 10^{14}$ M$_\odot$ per year, and 
several hundred under more realistic assumptions, many beyond redshift one. 
\item The cluster sample will constrain a combination of key cosmological 
parameters and the process of structure formation. The cluster survey 
will permit optical and 
X-ray follow-up to measure individual parameters, such as 
the mean matter density, the power spectrum amplitude, and gas density and 
temperature structures and their evolution, to good approximation. 
Selecting clusters from a deep SZ survey also provides 
for a very efficient use of X-ray observation time. 
Those at high redshift will be key targets for 
multi-waveband studies of cluster evolution. 
\item Because the survey method provides a sample of high redshift 
clusters that is essentially selected by mass it will be possible to 
make reliable distance estimates at high redshift and hence measure 
$q_0$. 
\item Such an instrument will also be able to make highly detailed 
pointed observations of clusters. 
\end{enumerate}

\section*{Acknowledgments}

RK acknowledges support from an EU Marie Curie Fellowship. 
VRE and CG acknowledge support from PPARC Postdoctoral Fellowships.


\bsp 

\label{lastpage}

\end{document}